Negative parity tetraquarks with the open charm.

Gerasyuta S.M.[1,2] , Kochkin V.I.[1]

1. Department of Theoretical Physics, St. Petersburg State University, 198904, St. Petersburg, Russia.
2. Department of Physics, LTA, 194021, St. Petersburg, Russia,
   E-mail: gerasyuta@sg6488.spb.edu

Abstract

The relativistic four-quark equations are found in the framework of coupled-channel formalism. The dynamical mixing of the meson-meson states with the four-quark states is considered. The four-quark amplitudes of the negative parity tetraquarks including the quarks of three flavors (u, d, s) and the charmed quark are constructed. The poles of these amplitudes determine the masses of tetraquarks. The mass values of low-lying tetraquarks with the spin-parity $J^P = 0^-, 1^-, 2^-, 3^-$ are calculated.

PACS number: 11.55Fv, 12.39 Ki, 12.39Mk, 12.40 Yx

## I. Introduction.

In 2006, BaBar Collaboration observed a new $c\bar{s}$ state $D_{SJ}^*(2860)$ [1] with a mass M=2856 MeV and width $\Gamma$=48 MeV. At the same time, Belle Collaboration reported a broader $c\bar{s}$ state $D_{S1}^*(2700)$ with $J^P=1^-$ in $B^+ \to \bar{D}^0 D^0 K^+$ decay [2]. Its mass is M=2710 MeV and width $\Gamma$=125 MeV.

More recently, the $D_{SJ}^*(2860)$ and $D_{S1}^*(2700)$ were found again by BaBar Collaboration in both $DK$ and $D^*K$ channels [3]. The observation of the $D_{SJ}^*(2860)$ in both $DK$ and $D^*K$ channels rules out it to be the $0^+$ state, since a $^3P_0$ $c\bar{s}$ state is forbidden to decay into $D^*K$. Its possible spin-parity should be $J^P=1^-,2^+,3^-$.

$D_{SJ}^*(2860)$ was proposed as the first radial excitation of $D_{SJ}^*(2317)$ in Ref. [4], as a $J^P=3^-$ $c\bar{s}$ state in Ref. [5] and as 2P $c\bar{s}$ state in Ref. [6]. Apart from the $D_{SJ}^*(2860)$ and $D_{S1}^*(2700)$ in the $D^*K$ channel the BaBar Collaboration also found the evidence for the $D_{SJ}^*(3040)$, whose mass and decay properties have been discussed in Refs. [7-9]. According to the heavy quark effective field theory, heavy mesons form doublets. For example, we have one S-wave doublet $(0^-,1^-)=(D_S(1965), D_S^*(2115))$ and two P-wave doublets $(0^+,1^+)=(D_{SJ}(2460), D_{SJ}^*(2317))$ and $(1^+,2^+)=(D_{S1}(2536), D_{SJ}^*(2573))$. The two D-wave $c\bar{s}$ doublets $(1^-,2^-)$ and $(2^-,3^-)$ have not been observed yet.

Systematic studies of the heavy-light meson spectra in quark models show that the expected $D_S$ masses are about 2.66-2.80 GeV for $1^-[D_S(2^3S_1)]$, 2.70-2.90 GeV for $1^-[D_S(1^3D_1)]$, 2.80-3.00 GeV for $3^-[D_S(2^3D_3)]$, 3.00-3.20 GeV for $2^+[D_S(2^3P_2)]$ and 3.10-3.30 GeV for $2^+[D_S(1^3F_2)]$, respectively [10-17].

The $J^P=1^-$ assignment for the $D_{S1}^*(2700)$ is consistent with the experiment. It would be most plausible that the $D_{SJ}^*(2860)$ is assigned to be either a $1^-$ or $3^-$ state. In the pure S-wave or D-wave $c\bar{s}$ picture, the decay properties of the $D_{S1}^*(2700)$ and $D_{SJ}^*(2860)$ have been studied for several different quantum numbers using various approaches [18-22].

In our papers [23, 24] relativistic of three-body Faddeev equations was obtained in the form of dispersion relations in the pair energy of two interacting particles. The mass spectra of S-wave baryons including $u, d, s, c$ quarks were calculated by a method based on isolating the leading singularities in the amplitude. We searched for the approximate solution of integral three-quark equation by taking into account two-particle and triangle singularities, all the weaker ones being neglected. If we considered such an approximation, which corresponds to taking into account two-body and triangle singularities, and defined all the smooth functions in the middle point the physical region of the Dalitz plot, then the problem was reduced to solving a system of simple algebraic equations.

In the previous paper [25] the relativistic four-quark equations are found in the framework of coupled-channel formalism. The dynamical mixing of the meson-meson states with the four-quark states is considered. The four-quark amplitudes of cryptoexotic mesons including the quarks of three flavors $u, d, s$ and the charmed quark are constructed. The low-lying meson-meson state masses and the contributions of subamplitudes to the tetraquark amplitudes are calculated.

The present paper is organized as follows. In Sec. II, we obtain the relativistic four-particle equations which describe the interaction of quarks for the excited tetraquarks with the charm. In Sec. III the coupled equations for the reduced amplitudes are derived. Section IV is devoted to a calculation of the mass spectrum of negative parity tetraquarks with the open charm (Tables I-IV). In the conclusion, the status of the considered model is discussed.

II. Four-quark amplitudes for the negative parity tetraquarks.

We derive the relativistic four-quark equations in the framework of the dispersion relation technique. We use only planar diagrams; the other diagrams due to the rules of $1/N_c$ expansion [26-28] are neglected. The current generates a four-quark system. Their successive pair interactions lead to the diagrams shown in Fig. 1. The correct equations for the amplitude are obtained by taking into account all possible subamplitudes. It corresponds to the division of complete system into subsystems with smaller number of particles. Then one should represent a four-particle amplitude as a sum of six subamplitudes:

$$A = A_{12} + A_{13} + A_{14} + A_{23} + A_{24} + A_{34}. \tag{1}$$

This defines the division of the diagrams into groups according to the certain pair interaction of particles. The total amplitude can be represented graphically as a sum of diagrams. We need to consider only one group of diagrams and the amplitude corresponding to them, for example $A_{12}$. The derivation of the relativistic generalization of the Faddeev-Yakubovsky approach for the tetraquark is considered. We shall construct the four-quark amplitude of $c\bar{s}u\bar{u}$ meson. The set of diagrams associated with the amplitude $A_{12}$ can further be broken down into groups corresponding to subamplitudes: $A_1(s, s_{12}, s_{34})$, $A_2(s, s_{23}, s_{14})$, $A_3(s, s_{12}, s_{34})$, $A_4(s, s_{23}, s_{14})$, $A_5(s, s_{23}, s_{123})$, $A_6(s, s_{14}, s_{134})$, $A_7(s, s_{34}, s_{234})$, $A_8(s, s_{12}, s_{123})$ (Fig. 1), if we consider the tetraquark with $J^P = 3^-$. Here $s_{ik}$ is the two-particle subenergy squared, $s_{ijk}$ corresponds to the energy squared of particles $i$, $j$, $k$ and $s$ is the system total energy squared.

The system of graphical equations is determined by the subamplitudes using the self-consistent method. The coefficients are determined by the permutation of quarks [29, 30].

In order to represent the subamplitudes $A_1(s, s_{12}, s_{34})$, $A_2(s, s_{23}, s_{14})$, $A_3(s, s_{12}, s_{34})$, $A_4(s, s_{23}, s_{14})$, $A_5(s, s_{23}, s_{123})$, $A_6(s, s_{14}, s_{134})$, $A_7(s, s_{34}, s_{234})$, $A_8(s, s_{12}, s_{123})$ in the form of a dispersion relation it is necessary to define the amplitudes of quark-antiquark interaction $a_n(s_{ik})$. The pair quarks amplitudes $q\bar{q} \to q\bar{q}$ are calculated in the framework of the dispersion N/D method with the input four-fermion interaction [31, 32] with the quantum numbers of the gluon [33]. The regularization of the dispersion integral for the D-function is carried out with the cutoff parameters $\Lambda$.

The four-quark interaction is considered as an input:

$$g_v(\bar{q}\lambda I_f \gamma_\mu q)^2 + 2g_v^{(s)}(\bar{q}\lambda\gamma_\mu I_f q)(\bar{s}\lambda\gamma_\mu s) + g_v^{(ss)}(\bar{s}\lambda\gamma_\mu s)^2 + \\ + 2g_v^{(c)}(\bar{q}\lambda\gamma_\mu I_f q)(\bar{c}\lambda\gamma_\mu c) + 2g_v^{(cs)}(\bar{s}\lambda\gamma_\mu s)(\bar{c}\lambda\gamma_\mu c) \tag{2}$$

Here $I_f$ is the unity matrix in the flavor space (u, d), $\lambda$ are the color Gell-Mann matrices. In order to avoid additional violation parameters we introduce the scale shift of the dimensional parameters [33]:

$$g = \frac{m^2}{\pi^2} g_v = \frac{(m+m_s)^2}{4\pi^2} g_v^{(s)} = \frac{m_s^2}{\pi^2} g_v^{(ss)} = \frac{(m+m_c)^2}{4\pi^2} g_v^{(c)} = \frac{(m_s+m_c)^2}{4\pi^2} g_v^{(cs)}, \quad (3)$$

$$\Lambda = \frac{4\Lambda(ik)}{(m_i+m_k)^2}. \quad (4)$$

Here $m_i$ and $m_k$ are the quark masses in the intermediate state of the quark loop. Dimensionless parameters g and $\Lambda$ are supposed to be constants which are independent of the quark interaction type. The applicability of Eq. (2) is verified by the success of De Rujula-Georgi-Glashow quark model [34], where only the short-range part of Breit potential connected with the gluon exchange is responsible for the mass splitting in hadron multiplets.

We use the results of our relativistic quark model [33] and write down the pair quark amplitudes in the form:

$$a_n(s_{ik}) = \frac{G_n^2(s_{ik})}{1 - B_n(s_{ik})}, \quad (5)$$

$$B_n(s_{ik}) = \int_{(m_i+m_k)^2}^{(m_i+m_k)^2 \Lambda/4} \frac{ds'_{ik}}{\pi} \frac{\rho_n(s'_{ik}) G_n^2(s'_{ik})}{s'_{ik} - s_{ik}}. \quad (6)$$

Here $G_n(s_{ik})$ are the quark-antiquark vertex functions (Table V, VI). The vertex functions are determined by the contribution of the crossing channels. The vertex functions satisfy the Fierz relations. All of these vertex functions are generated from $g_v$, $g_v^{(s)}$, $g_v^{(ss)}$, $g_v^{(c)}$ and $g_v^{(cs)}$. $B_n(s_{ik})$ and $\rho_n(s_{ik})$ are the Chew-Mandelstam functions with cutoff $\Lambda$ [35] and the phase space, respectively:

$$\rho_n(s_{ik}) = \left( \alpha(n) \frac{s_{ik}}{(m_i+m_k)^2} + \beta(n) + \delta(n) \frac{(m_i-m_k)^2}{s_{ik}} \right) \times$$
$$\times \frac{\sqrt{[s_{ik} - (m_i+m_k)^2][s_{ik} - (m_i-m_k)^2]}}{s_{ik}}, \quad (7)$$

The coefficients $\alpha(n)$, $\beta(n)$ and $\delta(n)$ are given in Table VII, VIII.

Here n corresponds to a $q\bar{q}$ pair in the color state $1_c$. In the case in question, the interacting quarks do not produce a bound state, therefore the integration in Eqs. (8) - (15) is carried out from the threshold $(m_i + m_k)^2$ to the cutoff $\Lambda(ik)$. The coupled integral equation systems, corresponding to Fig. 1 (the meson state with n=5 and $J^{PC} = 3^-$ for the $c\bar{s}u\bar{u}$) can be described as:

$$A_1(s, s_{12}, s_{34}) = \frac{\lambda_1 B_4(s_{12}) B_1(s_{34})}{[1 - B_4(s_{12})][1 - B_1(s_{34})]} + 2\hat{J}_2(s_{12}, s_{34}, 4, 1) A_5(s, s'_{23}, s'_{123}) + \\ + 2\hat{J}_2(s_{12}, s_{34}, 4, 1) A_6(s, s'_{14}, s'_{134})$$
(8)

$$A_2(s, s_{23}, s_{14}) = \frac{\lambda_2 B_1(s_{23}) B_4(s_{14})}{[1 - B_1(s_{23})][1 - B_4(s_{14})]} + 2\hat{J}_2(s_{23}, s_{14}, 1, 4) A_7(s, s'_{34}, s'_{234}) + \\ + 2\hat{J}_2(s_{23}, s_{14}, 1, 4) A_8(s, s'_{12}, s'_{123})$$
(9)

$$A_3(s, s_{12}, s_{34}) = \frac{\lambda_3 B_1(s_{12}) B_4(s_{34})}{[1 - B_1(s_{12})][1 - B_4(s_{34})]} + 2\hat{J}_2(s_{12}, s_{34}, 1, 4) A_5(s, s'_{23}, s'_{123}) + \\ + 2\hat{J}_2(s_{12}, s_{34}, 1, 4) A_6(s, s'_{14}, s'_{134})$$
(10)

$$A_4(s, s_{23}, s_{14}) = \frac{\lambda_4 B_4(s_{23}) B_1(s_{14})}{[1 - B_4(s_{23})][1 - B_1(s_{14})]} + 2\hat{J}_2(s_{23}, s_{14}, 4, 1) A_7(s, s'_{34}, s'_{234}) + \\ + 2\hat{J}_2(s_{23}, s_{14}, 4, 1) A_8(s, s'_{12}, s'_{123})$$
(11)

$$A_5(s, s_{23}, s_{123}) = \frac{\lambda_5 B_5(s_{23})}{1 - B_5(s_{23})} + 2\hat{J}_3(s_{23}, 5) A_1(s, s'_{12}, s'_{34}) + 2\hat{J}_3(s_{23}, 5) A_3(s, s'_{12}, s'_{34}) + \\ + \hat{J}_1(s_{23}, 5) A_7(s, s'_{34}, s'_{234}) + \hat{J}_1(s_{23}, 5) A_8(s, s'_{12}, s'_{123})$$
(12)

$$A_6(s, s_{14}, s_{134}) = \frac{\lambda_6 B_5(s_{14})}{1 - B_5(s_{14})} + 2\hat{J}_3(s_{14}, 5) A_1(s, s'_{12}, s'_{34}) + 2\hat{J}_3(s_{14}, 5) A_3(s, s'_{12}, s'_{34}) + \\ + \hat{J}_1(s_{14}, 5) A_7(s, s'_{34}, s'_{234}) + \hat{J}_1(s_{14}, 5) A_8(s, s'_{12}, s'_{123})$$
(13)

$$A_7(s, s_{34}, s_{234}) = \frac{\lambda_7 B_5(s_{34})}{1 - B_5(s_{34})} + 2\hat{J}_3(s_{34}, 5) A_4(s, s'_{23}, s'_{14}) + 2\hat{J}_3(s_{34}, 5) A_2(s, s'_{23}, s'_{14}) + \\ + \hat{J}_1(s_{34}, 5) A_5(s, s'_{23}, s'_{123}) + \hat{J}_1(s_{34}, 5) A_6(s, s'_{14}, s'_{134})$$
(14)

$$A_8(s, s_{12}, s_{123}) = \frac{\lambda_8 B_5(s_{12})}{1 - B_5(s_{12})} + 2\hat{J}_3(s_{12}, 5) A_4(s, s'_{23}, s'_{14}) + 2\hat{J}_3(s_{12}, 5) A_2(s, s'_{23}, s'_{14}) + \\ + \hat{J}_1(s_{12}, 5) A_6(s, s'_{14}, s'_{134}) + \hat{J}_1(s_{12}, 5) A_5(s, s'_{23}, s'_{123})$$
(15)

where $\lambda_i$ are the current constants. They do not affect the mass spectrum of tetraquarks. We introduce the integral operators:

$$\hat{J}_1(s_{12},l) = \frac{G_l(s_{12})}{[1-B_l(s_{12})]} \int\limits_{(m_1+m_2)^2}^{(m_1+m_2)^2\Lambda/4} \frac{ds'_{12}}{\pi} \frac{G_l(s'_{12})\rho_l(s'_{12})}{s'_{12}-s_{12}} \int\limits_{-1}^{+1} \frac{dz_1}{2}, \tag{16}$$

$$\hat{J}_2(s_{12},s_{34},l,p) = \frac{G_l(s_{12})G_p(s_{34})}{[1-B_l(s_{12})][1-B_p(s_{34})]} \times$$
$$\times \int\limits_{(m_1+m_2)^2}^{(m_1+m_2)^2\Lambda/4} \frac{ds'_{12}}{\pi} \frac{G_l(s'_{12})\rho_l(s'_{12})}{s'_{12}-s_{12}} \int\limits_{(m_3+m_4)^2}^{(m_3+m_4)^2\Lambda/4} \frac{ds'_{34}}{\pi} \frac{G_p(s'_{34})\rho_p(s'_{34})}{s'_{34}-s_{34}} \int\limits_{-1}^{+1} \frac{dz_3}{2} \int\limits_{-1}^{+1} \frac{dz_4}{2}, \tag{17}$$

$$\hat{J}_3(s_{12},l) = \frac{G_l(s_{12},\widetilde{\Lambda})}{1-B_l(s_{12},\widetilde{\Lambda})} \times$$
$$\times \frac{1}{4\pi} \int\limits_{(m_1+m_2)^2}^{(m_1+m_2)^2\widetilde{\Lambda}/4} \frac{ds'_{12}}{\pi} \frac{G_l(s'_{12},\widetilde{\Lambda})\rho_l(s'_{12})}{s'_{12}-s_{12}} \int\limits_{-1}^{+1} \frac{dz_1}{2} \int\limits_{-1}^{+1} dz \int\limits_{z_2^-}^{z_2^+} dz_2 \frac{1}{\sqrt{1-z^2-z_1^2-z_2^2+2zz_1z_2}}, \tag{18}$$

where $l, p$ are equal to 1 - 11. Here $m_i$ is a quark mass.

In Eqs. (16) and (18) $z_1$ is the cosine of the angle between the relative momentum of the particles 1 and 2 in the intermediate state and the momentum of the particle 3 in the final state, taken in the c.m. of particles 1 and 2. In Eq. (18) $z$ is the cosine of the angle between the momenta of the particles 3 and 4 in the final state, taken in the c.m. of particles 1 and 2. $z_2$ is the cosine of the angle between the relative momentum of particles 1 and 2 in the intermediate state and the momentum of the particle 4 in the final state, is taken in the c.m. of particles 1 and 2. In Eq. (17): $z_3$ is the cosine of the angle between relative momentum of particles 1 and 2 in the intermediate state and the relative momentum of particles 3 and 4 in the intermediate state, taken in the c.m. of particles 1 and 2. $z_4$ is the cosine of the angle between the relative momentum of the particles 3 and 4 in the intermediate state and that of the momentum of the particle 1 in the intermediate state, taken in the c.m. of particles 3, 4.

We can pass from the integration over the cosines of the angles to the integration over the subenergies [36].

Let us extract two-particle singularities in the amplitudes $A_1(s,s_{12},s_{34})$, $A_2(s,s_{23},s_{14})$, $A_3(s,s_{12},s_{34})$, $A_4(s,s_{23},s_{14})$, $A_5(s,s_{23},s_{123})$, $A_6(s,s_{14},s_{134})$, $A_7(s,s_{34},s_{234})$, $A_8(s,s_{12},s_{123})$:

$$A_j(s,s_{ik},s_{lm}) = \frac{\alpha_j(s,s_{ik},s_{lm})B_1(s_{ik})B_4(s_{lm})}{[1-B_1(s_{ik})][1-B_4(s_{lm})]}, \qquad j=1\text{-}4, \tag{19}$$

$$A_j(s, s_{ik}, s_{ikl}) = \frac{\alpha_j(s, s_{ik}, s_{ikl}) B_5(s_{ik})}{1 - B_5(s_{ik})}, \qquad j=5\text{-}8, \qquad (20)$$

We do not extract three-particle singularities, because they are weaker than two-particle singularities.

We used the classification of singularities, which was proposed in paper [37]. The construction of approximate solution of Eqs. (8) - (15) is based on the extraction of the leading singularities of the amplitudes. The main singularities in $s_{ik} \approx (m_i + m_k)^2$ are from pair rescattering of the particles i and k. First of all there are threshold square-root singularities. Also possible are pole singularities which correspond to the bound states. The diagrams of Fig.1 apart from two-particle singularities have the triangular singularities and the singularities defining the interaction of four particles. Such classification allows us to search the corresponding solution of Eqs. (8) - (15) by taking into account some definite number of leading singularities and neglecting all the weaker ones. We consider the approximation which defines two-particle, triangle and four-particle singularities. The functions $\alpha_1(s, s_{12}, s_{34})$, $\alpha_2(s, s_{23}, s_{14})$, $\alpha_3(s, s_{12}, s_{34})$, $\alpha_4(s, s_{23}, s_{14})$, $\alpha_5(s, s_{23}, s_{123})$, $\alpha_6(s, s_{14}, s_{134})$, $\alpha_7(s, s_{34}, s_{234})$, $\alpha_8(s, s_{12}, s_{123})$ are the smooth functions of $s_{ik}$, $s_{ikl}$, $s$ as compared with the singular part of the amplitudes, hence they can be expanded in a series in the singularity point and only the first term of this series should be employed further. Using this classification, one defines the reduced amplitudes $\alpha_1$, $\alpha_2$, $\alpha_3$, $\alpha_4$, $\alpha_5$, $\alpha_6$, $\alpha_7$, $\alpha_8$ as well as the B-functions in the middle point of the physical region of Dalitz-plot at the point $s_0$:

$$s_0^{ik} = 0.25 (m_i + m_k)^2 s_0 \qquad (21)$$

$$s_{123} = 0.25 s_0 \sum_{\substack{i,k=1 \\ i \neq k}}^{3} (m_i + m_k)^2 - \sum_{i=1}^{3} m_i^2, \quad s_0 = \frac{s + 2\sum_{i=1}^{4} m_i^2}{0.25 \sum_{\substack{i,k=1 \\ i \neq k}}^{4} (m_i + m_k)^2}$$

Such a choice of point $s_0$ allows us to replace the integral Eqs. (8) - (15) (Fig. 1) by the algebraic equations (22) - (29) respectively:

$$\alpha_1 = \lambda_1 + 2\alpha_5 JB_1(4,1,5) + 2\alpha_6 JB_2(4,1,5), \qquad (22)$$

$$\alpha_2 = \lambda_2 + 2\alpha_7 JB_3(1,4,5) + 2\alpha_8 JB_4(1,4,5), \qquad (23)$$

$$\alpha_3 = \lambda_3 + 2\alpha_5 JB_5(1,4,5) + 2\alpha_6 JB_6(1,4,5), \tag{24}$$

$$\alpha_4 = \lambda_4 + 2\alpha_7 JB_7(4,1,5) + 2\alpha_8 JB_8(4,1,5), \tag{25}$$

$$\alpha_5 = \lambda_5 + 2\alpha_1 JC_1(5,4,1) + 2\alpha_3 JC_2(5,1,4) + \alpha_7 JA_1(5) + \alpha_8 JA_2(5), \tag{26}$$

$$\alpha_6 = \lambda_6 + 2\alpha_1 JC_3(5,4,1) + 2\alpha_3 JC_4(5,1,4) + \alpha_7 JA_3(5) + \alpha_8 JA_4(5), \tag{27}$$

$$\alpha_7 = \lambda_7 + 2\alpha_4 JC_5(5,4,1) + 2\alpha_2 JC_6(5,1,4) + \alpha_5 JA_5(5) + \alpha_6 JA_6(5), \tag{28}$$

$$\alpha_8 = \lambda_8 + 2\alpha_4 JC_7(5,4,1) + 2\alpha_2 JC_8(5,1,4) + \alpha_6 JA_7(5) + \alpha_5 JA_8(5), \tag{29}$$

We use the functions $JA_i(l)$, $JB_i(l,p,r)$, $JC_i(l,p,r)$ ($l,p,r = 1 - 11$), which are determined by the various $s_0^{ik}$ (Eq. 21). These functions are similar to the functions:

$$JA_4(l) = \frac{G_l^2(s_0^{12})B_l(s_0^{23})}{B_l(s_0^{12})} \int\limits_{(m_1+m_2)^2}^{(m_1+m_2)^2 \Lambda/4} \frac{ds'_{12}}{\pi} \frac{\rho_l(s'_{12})}{s'_{12} - s_0^{12}} \int\limits_{-1}^{+1} \frac{dz_1}{2} \frac{1}{1 - B_l(s'_{23})}, \tag{30}$$

$$JB_1(l,p,r) = \frac{G_l^2(s_0^{12})G_p^2(s_0^{34})B_r(s_0^{23})}{B_l(s_0^{12})B_p(s_0^{34})} \times$$

$$\times \int\limits_{(m_1+m_2)^2}^{(m_1+m_2)^2 \Lambda/4} \frac{ds'_{12}}{\pi} \frac{\rho_l(s'_{12})}{s'_{12} - s_0^{12}} \int\limits_{(m_3+m_4)^2}^{(m_3+m_4)^2 \Lambda/4} \frac{ds'_{34}}{\pi} \frac{\rho_p(s'_{34})}{s'_{34} - s_0^{34}} \int\limits_{-1}^{+1} \frac{dz_3}{2} \int\limits_{-1}^{+1} \frac{dz_4}{2} \frac{1}{1 - B_r(s'_{23})} \tag{31}$$

$$JC_3(l,p,r) = \frac{G_l^2(s_0^{12}, \widetilde{\Lambda})B_p(s_0^{23})B_r(s_0^{14})}{1 - B_l(s_0^{12}, \widetilde{\Lambda})} \frac{1 - B_l(s_0^{12})}{B_l(s_0^{12})} \times$$

$$\times \frac{1}{4\pi} \int\limits_{(m_1+m_2)^2}^{(m_1+m_2)^2 \widetilde{\Lambda}/4} \frac{ds'_{12}}{\pi} \frac{\rho_l(s'_{12})}{s'_{12} - s_0^{12}} \int\limits_{-1}^{+1} \frac{dz_1}{2} \int\limits_{-1}^{+1} dz \int\limits_{z_2^-}^{z_2^+} dz_2 \frac{1}{\sqrt{1 - z^2 - z_1^2 - z_2^2 + 2zz_1z_2}} \times \tag{32}$$

$$\times \frac{1}{[1 - B_p(s'_{23})][1 - B_r(s'_{14})]}$$

$$\widetilde{\Lambda}(ik) = \begin{cases} \Lambda(ik), \text{ if } \Lambda(ik) \leq (\sqrt{s_{123}} + m_3)^2 \\ (\sqrt{s_{123}} + m_3)^2, \text{ if } \Lambda(ik) > (\sqrt{s_{123}} + m_3)^2 \end{cases} \tag{33}$$

The other choices of point $s_0$ do not change essentially the contributions of $\alpha_1$, $\alpha_2$, $\alpha_3$, $\alpha_4$, $\alpha_5$, $\alpha_6$, $\alpha_7$, $\alpha_8$ therefore we omit the indices $s_0^{ik}$. Since the vertex functions depend only slightly on energy it is possible to treat them as constants in our approximation.

The solutions of the system of equations are considered as:

$$\alpha_i(s) = F_i(s,\lambda_i)/D(s), \qquad (34)$$

where zeros of $D(s)$ determinants define the masses of bound states of tetraquarks. $F_i(s,\lambda_i)$ determine the contributions of subamplitudes for the tetraquark amplitude.

### III. Calculation results.

The pole of the reduced amplitudes $\alpha_i$ corresponds to the bound state and determines the mass of the meson-meson state with n=5 and $J^{PC}=3^-$ for the $c\bar{s}u\bar{u}$ (Fig. 1). The quark masses of model $m_{u,d}$=385 MeV, $m_s$=510 MeV coincide with the ordinary meson model ones [33]. In order to fix $m_c$=1586 MeV, the tetraquark mass for the $J^{PC}=2^{++}$ X(3940) is taken into account. The present model do not use the new parameters. The parameters are determined by fixing the masses for the $J^{PC}=1^{++}$ X(3872) and $J^{PC}=2^{++}$ X(3940). The model in question has only two parameters, the cutoff $\Lambda$=10, and the gluon coupling constant $g$=0.794. The parameters are similar to the Ref. [25] ones. This relativistic four-body approach gives rise to the dynamical mixing of the meson-meson states with the four-quark states. The masses of meson-meson states with isospin I=0, ½ and the spin-parity $J^P=0^-,1^-,2^-,3^-$ are predicted (Tables I-IV). We calculated the open charmed tetraquark M=2758 MeV with the spin-parity $J^P=3^-$ and the quark content $c\bar{s}u\bar{u}$. It is similar to the experimental value M=2860 MeV [38]. The second result of our model was obtained for the tetraquark with the mass M=2696 MeV spin-parity $J^P=1^-$ and the quark content $c\bar{s}u\bar{u}$ (the experimental value is M=2710 MeV). The contributions of the subamplitudes to the tetraquark amplitude are given in the Table I. The contributions of the meson-meson subamplitudes are about 30-50 percent.

IV. Conclusion.

In a strongly bound system of light and heavy quarks, such as the tetraquarks, the approximation of nonrelativistic kinematics and dynamics is not justified. The relativistic quark model for the study of the heavy meson spectroscopy is constructed using the dispersion N/D method.

In our previous paper [25] we showed that the contributions of the tetraquarks to the ordinary mesons with the spin-parity $J^P = 0^+$ are about 5-10 percent, therefore the widths of these states are determined by the ordinary state widths. In the present paper we considered the low-lying meson-meson states with the quark contents: $(c\bar{s})(u\bar{u})$, $(c\bar{u})(u\bar{u})$, and $(c\bar{s})(s\bar{s})$. We calculated the masses and the contributions of subamplitudes to the negative parity tetraquark amplitudes [Table I-IV]. But the two D-wave $c\bar{s}$ doublets (1-,2-) and (2-,3-) have not been observed yet. Therefore we cannot consider the mixing of the negative parity tetraquarks with the ordinary states $c\bar{s}$. We considered the tetraquarks with the spin-parity $J^P = 0^-, 1^-, 2^-, 3^-$. The calculated masses of tetraquarks with the quark content $(c\bar{s})(u\bar{u})$ in the Table I are given $J^P = 0^-, 1^-$ ($M_{0^-}$=2743 MeV, $M_{1^-}$=2500 MeV), $J^P = 1^-, 2^-$ ($M_{1^-}$=2696 MeV, $M_{2^-}$=2750 MeV), $J^P = 2^-, 3^-$ ($M_{2^-}$=2495 MeV, $M_{3^-}$=2758 MeV).

The strongly decays of the $D^*_{S1}(2700)$ and $D^*_{SJ}(2860)$ are investigated in the framework of the $^3P_0$ model [39]. Compare the predicted mass from the model and strong decay properties from $^3P_0$ model, we believe that the further experimental information on the $D^*_{S1}(2700)$ and $D^*_{SJ}(2860)$ is needed.

Acknowledgments

The authors would like to thank T. Barnes, E. Oset and S.L. Zhu for useful discussions. The work was carried with the support of the Russian Ministry of Education (grant 2.1.1.68.26).

Table I. Low-lying meson-meson state masses (MeV) and the contributions of subamplitudes to the tetraquark amplitudes (in percent) for the $J^P = 3^-$, n=5.

| $(c\bar{s})(u\bar{u})$ | $J^P = 3^-$ | $(c\bar{u})(u\bar{u})$ | $J^P = 3^-$ | $(c\bar{s})(s\bar{s})$ | $J^P = 3^-$ |
|---|---|---|---|---|---|
| Masses: | 2758 MeV | Masses: | 2663 MeV | Masses: | 2956 MeV |
| $(c\bar{s})_{2^+}(u\bar{u})_{1^-}$ | 10.33 | $(c\bar{u})_{2^+}(u\bar{u})_{1^-}$ | 18.18 | $(c\bar{s})_{2^+}(s\bar{s})_{1^-}$ | 18.76 |
| $(c\bar{u})_{2^+}(u\bar{s})_{1^-}$ | 8.34 | $(c\bar{u})_{1^-}(u\bar{u})_{2^+}$ | 18.09 | $(c\bar{s})_{1^-}(s\bar{s})_{2^+}$ | 18.70 |
| $(c\bar{s})_{1^-}(u\bar{u})_{2^+}$ | 10.29 | | | | |
| $(c\bar{u})_{1^-}(u\bar{s})_{2^+}$ | 8.31 | | | | |

Table II.a. Low-lying meson-meson state masses (MeV) and the contributions of subamplitudes to the tetraquark amplitudes (in percent) for the $J^{PC} = 2^-$, n=6.

| $(c\bar{s})(u\bar{u})$ | $J^{PC} = 2^-$ | $(c\bar{u})(u\bar{u})$ | $J^{PC} = 2^-$ | $(c\bar{s})(s\bar{s})$ | $J^{PC} = 2^-$ |
|---|---|---|---|---|---|
| Masses: | 2495 MeV | Masses: | 2423 MeV | Masses: | 2655 MeV |
| $(c\bar{s})_{1^-}(u\bar{u})_{1^+}$ | 6.35 | $(c\bar{u})_{1^-}(u\bar{u})_{1^+}$ | 12.42 | $(c\bar{s})_{1^-}(s\bar{s})_{1^+}$ | 12.63 |
| $(c\bar{s})_{1^+}(u\bar{u})_{1^-}$ | 6.26 | $(c\bar{u})_{1^+}(u\bar{u})_{1^-}$ | 12.19 | $(c\bar{s})_{1^+}(s\bar{s})_{1^-}$ | 12.47 |
| $(c\bar{u})_{1^-}(u\bar{s})_{1^+}$ | 5.04 | $(c\bar{u})_{1^-}(u\bar{u})_{2^+}$ | 9.74 | $(c\bar{s})_{1^-}(s\bar{s})_{2^+}$ | 10.09 |
| $(c\bar{u})_{1^+}(u\bar{s})_{1^-}$ | 5.01 | $(c\bar{u})_{2^+}(u\bar{u})_{1^-}$ | 9.77 | $(c\bar{s})_{2^+}(s\bar{s})_{1^-}$ | 10.11 |
| $(c\bar{s})_{1^-}(u\bar{u})_{2^+}$ | 4.98 | | | | |
| $(c\bar{s})_{2^+}(u\bar{u})_{1^-}$ | 5.00 | | | | |
| $(c\bar{u})_{1^-}(u\bar{s})_{2^+}$ | 4.03 | | | | |
| $(c\bar{u})_{2^+}(u\bar{s})_{1^-}$ | 4.04 | | | | |

Table II.b. Low-lying meson-meson state masses (MeV) and the contributions of subamplitudes to the tetraquark amplitudes (in percent) for the $J^{PC} = 2^-$, n=11.

| $(c\bar{s})(u\bar{u})$ | $J^{PC} = 2^-$ | $(c\bar{u})(u\bar{u})$ | $J^{PC} = 2^-$ | $(c\bar{s})(s\bar{s})$ | $J^{PC} = 2^-$ |
|---|---|---|---|---|---|
| Masses: | 2750 MeV | Masses: | 2667 MeV | Masses: | 2964 MeV |
| $(c\bar{s})_{2^+}(u\bar{u})_{0^-}$ | 16.01 | $(c\bar{u})_{2^+}(u\bar{u})_{0^-}$ | 16.95 | $(c\bar{s})_{2^+}(s\bar{s})_{0^-}$ | 17.78 |
| $(c\bar{u})_{2^+}(u\bar{s})_{0^-}$ | 2.00 | $(c\bar{u})_{0^-}(u\bar{u})_{2^+}$ | 14.55 | $(c\bar{s})_{0^-}(s\bar{s})_{2^+}$ | 16.12 |
| $(c\bar{s})_{0^-}(u\bar{u})_{2^+}$ | 14.41 | | | | |
| $(c\bar{u})_{0^-}(u\bar{s})_{2^+}$ | 1.75 | | | | |

Table III.a. Low-lying meson-meson state masses (MeV) and the contributions of subamplitudes to the tetraquark amplitudes (in percent) for the $J^{PC}=0^-$, n=7.

| $(c\bar{s})(u\bar{u})$ | $J^{PC}=0^-$ | $(c\bar{u})(u\bar{u})$ | $J^{PC}=0^-$ | $(c\bar{s})(s\bar{s})$ | $J^{PC}=0^-$ |
|---|---|---|---|---|---|
| Masses: | 2800 MeV | Masses: | 2710 MeV | Masses: | 3010 MeV |
| $(c\bar{s})_{1^+}(u\bar{u})_{1^-}$ | 13.61 | $(c\bar{u})_{1^+}(u\bar{u})_{1^-}$ | 24.39 | $(c\bar{s})_{1^+}(s\bar{s})_{1^-}$ | 24.19 |
| $(c\bar{u})_{1^+}(u\bar{s})_{1^-}$ | 10.04 | $(c\bar{u})_{1^-}(u\bar{u})_{1^+}$ | 25.17 | $(c\bar{s})_{1^-}(s\bar{s})_{1^+}$ | 24.69 |
| $(c\bar{s})_{1^-}(u\bar{u})_{1^+}$ | 13.92 | | | | |
| $(c\bar{u})_{1^-}(u\bar{s})_{1^+}$ | 10.68 | | | | |

Table III.b. Low-lying meson-meson state masses (MeV) and the contributions of subamplitudes to the tetraquark amplitudes (in percent) for the $J^{PC}=0^-$, n=8.

| $(c\bar{s})(u\bar{u})$ | $J^{PC}=0^-$ | $(c\bar{u})(u\bar{u})$ | $J^{PC}=0^-$ | $(c\bar{s})(s\bar{s})$ | $J^{PC}=0^-$ |
|---|---|---|---|---|---|
| Masses: | 2743 MeV | Masses: | 2662 MeV | Masses: | 2960 MeV |
| $(c\bar{s})_{1^+}(u\bar{u})_{1^-}$ | 14.64 | $(c\bar{u})_{1^+}(u\bar{u})_{1^-}$ | 13.00 | $(c\bar{s})_{1^+}(s\bar{s})_{1^-}$ | 14.03 |
| $(c\bar{u})_{1^+}(u\bar{s})_{1^-}$ | 1.88 | $(c\bar{u})_{1^-}(u\bar{u})_{1^+}$ | 14.89 | $(c\bar{s})_{1^-}(s\bar{s})_{1^+}$ | 15.40 |
| $(c\bar{s})_{1^-}(u\bar{u})_{1^+}$ | 16.18 | | | | |
| $(c\bar{u})_{1^-}(u\bar{s})_{1^+}$ | 2.13 | | | | |

Table IV.a Low-lying meson-meson state masses (MeV) and the contributions of subamplitudes to the tetraquark amplitudes (in percent) for the $J^{PC}=1^-$, n=1.

| $(c\bar{s})(u\bar{u})$ | $J^{PC}=1^-$ |
|---|---|
| Masses: | 2500 MeV |
| $(c\bar{s})_{1^-}(u\bar{u})_{1^+}$ | 5.48 |
| $(c\bar{s})_{1^+}(u\bar{u})_{1^-}$ | 5.41 |
| $(c\bar{u})_{1^-}(u\bar{s})_{1^+}$ | 4.27 |
| $(c\bar{u})_{1^+}(u\bar{s})_{1^-}$ | 4.25 |
| $(c\bar{s})_{1^-}(u\bar{u})_{2^+}$ | 4.31 |
| $(c\bar{s})_{2^+}(u\bar{u})_{1^-}$ | 4.32 |
| $(c\bar{u})_{1^-}(u\bar{s})_{2^+}$ | 3.42 |
| $(c\bar{u})_{2^+}(u\bar{s})_{1^-}$ | 3.43 |
| $(c\bar{s})_{1^-}(u\bar{u})_{0^+}$ | 3.83 |
| $(c\bar{s})_{0^+}(u\bar{u})_{1^-}$ | 3.88 |
| $(c\bar{u})_{1^-}(u\bar{s})_{0^+}$ | 3.09 |
| $(c\bar{u})_{0^+}(u\bar{s})_{1^-}$ | 4.31 |

Table IV.b Low-lying meson-meson state masses (MeV) and the contributions of subamplitudes to the tetraquark amplitudes (in percent) for the $J^{PC}=1^-$, n=10.

| $(c\bar{s})(u\bar{u})$ | $J^{PC}=1^-$ | $(c\bar{u})(u\bar{u})$ | $J^{PC}=1^-$ | $(c\bar{s})(s\bar{s})$ | $J^{PC}=1^-$ |
|---|---|---|---|---|---|
| Masses: | 2696 MeV | Masses: | 2621 MeV | Masses: | 2900 MeV |
| $(c\bar{s})_{1^+}(u\bar{u})_{0^-}$ | 16.94 | $(c\bar{u})_{1^+}(u\bar{u})_{0^-}$ | 17.71 | $(c\bar{s})_{1^+}(s\bar{s})_{0^-}$ | 18.37 |
| $(c\bar{u})_{1^+}(u\bar{s})_{0^-}$ | 1.73 | $(c\bar{u})_{0^-}(u\bar{u})_{1^+}$ | 15.99 | $(c\bar{s})_{0^-}(s\bar{s})_{1^+}$ | 17.22 |
| $(c\bar{s})_{0^-}(u\bar{u})_{1^+}$ | 15.79 | | | | |
| $(c\bar{u})_{0^-}(u\bar{s})_{1^+}$ | 1.59 | | | | |

Table V. Vertex functions for the table Ia-IVa.

| $J^{PC}$ | $G_n^2$ |
|---|---|
| $1^-$ (n=1) | $4g/3$ |
| $0^+$ (n=2) | $8g/3$ |
| $1^+$ (n=3) | $4g/3$ |
| $2^+$ (n=4) | $4g/3$ |
| $3^-$ (n=5) | $4g/3$ |
| $2^-$ (n=6) | $4g/3$ |
| $0^-$ (n=7) | $4g/3$ |

Table VI. Vertex functions for the table Ib-IVb.

| $J^{PC}$ | $G_n^2$ |
|---|---|
| $0^-$ (n=8) | $8g/3 - 4g(m_i + m_k)^2/(3s_{ik})$ |
| $1^+$ (n=9) | $8g/3 - 4g(m_i + m_k)^2/(3s_{ik})$ |
| $1^-$ (n=10) | $4g/3$ |
| $2^-$ (n=11) | $4g/3$ |

Table VII. Coefficients of Chew-Mandelstam functions for the table Ia-IVa.

| $J^{PC}$ | $\alpha$ | $\beta$ | $\delta$ |
|---|---|---|---|
| $1^-$ (n=1) | 1/3 | 1/6-e/3 | -1/6 |
| $0^+$ (n=2) | 1/2 | -1/2 | 0 |
| $1^+$ (n=3) | 1/2 | -e/2 | 0 |
| $2^+$ (n=4) | 3/10 | 1/5-3e/10 | -1/5 |
| $3^-$ (n=5) | 2/7 | 3/14-10e/7 | -3/14 |
| $2^-$ (n=6) | 4/7 | -1/14-3e/7 | 1/14 |
| $0^-$ (n=7) | 1 | -1 | 0 |

$e = (m_i - m_k)^2 / (m_i + m_k)^2$

Table VIII. Coefficients of Chew-Mandelstam functions for the table Ib-IVb.

| $J^{PC}$ | $\alpha$ | $\beta$ | $\delta$ |
|---|---|---|---|
| $0^-$ (n=8) | 1/2 | -e/2 | 0 |
| $1^+$ (n=9) | 1/2 | -e/2 | 0 |
| $1^-$ (n=10) | 2/3 | -e | 1/3 |
| $2^-$ (n=11) | 1/2 | -e/2 | 0 |

$e = (m_i - m_k)^2 / (m_i + m_k)^2$

Figure captions.

Fig.1. Graphic representation of the equations for the four-quark subamplitudes $A_k$ ($k$=1-8) in the case of n=5 and $J^{PC} = 3^-$ ($c\bar{s}u\bar{u}$).

References.

1. B. Aubert et al.,(BaBar Collaboration) Phys.Rev.Lett. 97,222001(2006)
2. K. Abe et al.,(Belle Collaboration) arXiv: 0608.031 [hep-ex].
3. B. Aubert et al.( BaBar Collaboration), Phys.Rev.D80,092003 (2009).
4. E.V. Beveren and G. Rupp, Phys.Rev.Lett. 97,202001 (2006).
5. P. Colangelo, F.D. Fazio and S. Nicotri, Phys.Lett.B642,48 (2006).
6. F.E. Close, C.E. Tomas, O. Lakhina and E.S. Swanson, Phys.Lett.B647,159 (2007).
7. B. Chen, D.X. Wang and A. Zhang, Phys. Rev. D80, 071502 (R) (2009).
8. Z.F. Sun and X. Liu, Phys. Rev. D80, 074037 (2009).
9. X.H. Zhong and Q. Zhao, Phys.Rev. D81,014031 (2010).
10. S. Godfrey and N. Isgur, Phys. Rev. D32, 189 (1985).
11. S.N. Gupta and J.M. Jonson, Phys. Rev. D51, 168 (1995).
12. J. Zeng, J.W. Van Orden and W. Roberts, Phys. Rev. D52, 5229 (1995).
13. D. Ebert, V.O. Galkin and R.N. Faustov, Phys. Rev. D57, 5663 (1998).
14. T.A. Lahde, C.J. Nyfalt and D.O. Riska, Nucl. Phys. A674, 141 (2000).

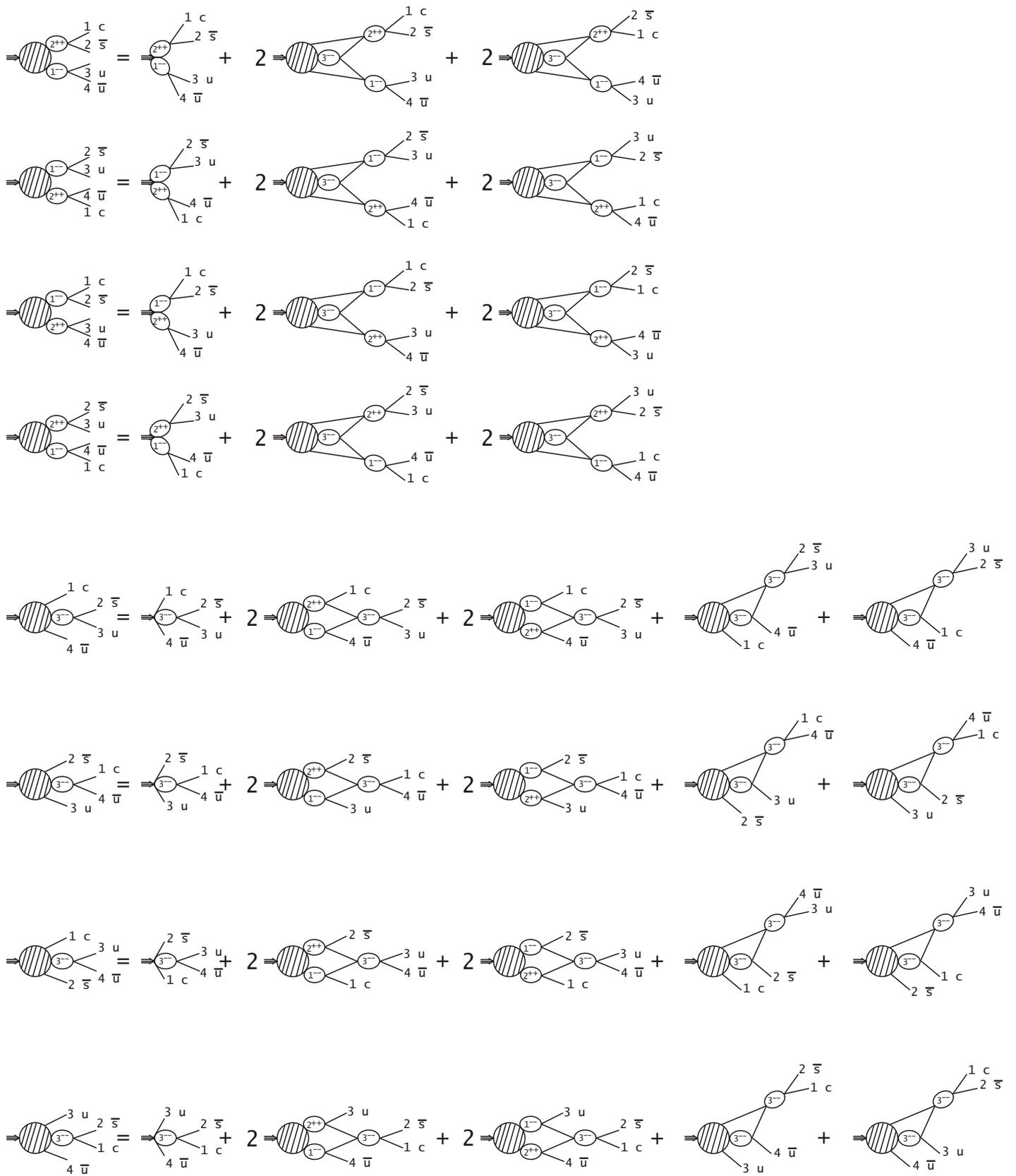

Fig. 1